\documentclass[twocolumn,showpacs,preprintnumbers,amsmath,amssymb,floatfix,prd,superscriptaddress]{revtex4}

\usepackage{graphicx}
\usepackage{dcolumn}
\usepackage{bm}
\usepackage{setspace}
\usepackage{revsymb}
\usepackage{url}
\usepackage{epsfig}

\newcommand{\met}{\mbox{$E_{T}\!\!\!\!\!\!\!/\,\,\,\,$}}

\newcommand{\rr}[2]{\raisebox{#2ex}[0pt]{#1}}

\begin{document}

\title{\large \bf \boldmath Search for new physics in the $\mu\mu+e/\mu+\met$ channel with a low-$p_T$ lepton threshold at the Collider Detector at Fermilab}

\affiliation{Institute of Physics, Academia Sinica, Taipei, Taiwan 11529, Republic of China} 
\affiliation{Argonne National Laboratory, Argonne, Illinois 60439} 
\affiliation{University of Athens, 157 71 Athens, Greece} 
\affiliation{Institut de Fisica d'Altes Energies, Universitat Autonoma de Barcelona, E-08193, Bellaterra (Barcelona), Spain} 
\affiliation{Baylor University, Waco, Texas  76798} 
\affiliation{Istituto Nazionale di Fisica Nucleare Bologna, $^v$University of Bologna, I-40127 Bologna, Italy} 
\affiliation{Brandeis University, Waltham, Massachusetts 02254} 
\affiliation{University of California, Davis, Davis, California  95616} 
\affiliation{University of California, Los Angeles, Los Angeles, California  90024} 
\affiliation{University of California, San Diego, La Jolla, California  92093} 
\affiliation{University of California, Santa Barbara, Santa Barbara, California 93106} 
\affiliation{Instituto de Fisica de Cantabria, CSIC-University of Cantabria, 39005 Santander, Spain} 
\affiliation{Carnegie Mellon University, Pittsburgh, PA  15213} 
\affiliation{Enrico Fermi Institute, University of Chicago, Chicago, Illinois 60637} 
\affiliation{Comenius University, 842 48 Bratislava, Slovakia; Institute of Experimental Physics, 040 01 Kosice, Slovakia} 
\affiliation{Joint Institute for Nuclear Research, RU-141980 Dubna, Russia} 
\affiliation{Duke University, Durham, North Carolina  27708} 
\affiliation{Fermi National Accelerator Laboratory, Batavia, Illinois 60510} 
\affiliation{University of Florida, Gainesville, Florida  32611} 
\affiliation{Laboratori Nazionali di Frascati, Istituto Nazionale di Fisica Nucleare, I-00044 Frascati, Italy} 
\affiliation{University of Geneva, CH-1211 Geneva 4, Switzerland} 
\affiliation{Glasgow University, Glasgow G12 8QQ, United Kingdom} 
\affiliation{Harvard University, Cambridge, Massachusetts 02138} 
\affiliation{Division of High Energy Physics, Department of Physics, University of Helsinki and Helsinki Institute of Physics, FIN-00014, Helsinki, Finland} 
\affiliation{University of Illinois, Urbana, Illinois 61801} 
\affiliation{The Johns Hopkins University, Baltimore, Maryland 21218} 
\affiliation{Institut f\"{u}r Experimentelle Kernphysik, Universit\"{a}t Karlsruhe, 76128 Karlsruhe, Germany} 
\affiliation{Center for High Energy Physics: Kyungpook National University, Daegu 702-701, Korea; Seoul National University, Seoul 151-742, Korea; Sungkyunkwan University, Suwon 440-746, Korea; Korea Institute of Science and Technology Information, Daejeon, 305-806, Korea; Chonnam National University, Gwangju, 500-757, Korea} 
\affiliation{Ernest Orlando Lawrence Berkeley National Laboratory, Berkeley, California 94720} 
\affiliation{University of Liverpool, Liverpool L69 7ZE, United Kingdom} 
\affiliation{University College London, London WC1E 6BT, United Kingdom} 
\affiliation{Centro de Investigaciones Energeticas Medioambientales y Tecnologicas, E-28040 Madrid, Spain} 
\affiliation{Massachusetts Institute of Technology, Cambridge, Massachusetts  02139} 
\affiliation{Institute of Particle Physics: McGill University, Montr\'{e}al, Qu\'{e}bec, Canada H3A~2T8; Simon Fraser University, Burnaby, British Columbia, Canada V5A~1S6; University of Toronto, Toronto, Ontario, Canada M5S~1A7; and TRIUMF, Vancouver, British Columbia, Canada V6T~2A3} 
\affiliation{University of Michigan, Ann Arbor, Michigan 48109} 
\affiliation{Michigan State University, East Lansing, Michigan  48824}
\affiliation{Institution for Theoretical and Experimental Physics, ITEP, Moscow 117259, Russia} 
\affiliation{University of New Mexico, Albuquerque, New Mexico 87131} 
\affiliation{Northwestern University, Evanston, Illinois  60208} 
\affiliation{The Ohio State University, Columbus, Ohio  43210} 
\affiliation{Okayama University, Okayama 700-8530, Japan} 
\affiliation{Osaka City University, Osaka 588, Japan} 
\affiliation{University of Oxford, Oxford OX1 3RH, United Kingdom} 
\affiliation{Istituto Nazionale di Fisica Nucleare, Sezione di Padova-Trento, $^w$University of Padova, I-35131 Padova, Italy} 
\affiliation{LPNHE, Universite Pierre et Marie Curie/IN2P3-CNRS, UMR7585, Paris, F-75252 France} 
\affiliation{University of Pennsylvania, Philadelphia, Pennsylvania 19104}
\affiliation{Istituto Nazionale di Fisica Nucleare Pisa, $^x$University of Pisa, $^y$University of Siena and $^z$Scuola Normale Superiore, I-56127 Pisa, Italy} 
\affiliation{University of Pittsburgh, Pittsburgh, Pennsylvania 15260} 
\affiliation{Purdue University, West Lafayette, Indiana 47907} 
\affiliation{University of Rochester, Rochester, New York 14627} 
\affiliation{The Rockefeller University, New York, New York 10021} 
\affiliation{Istituto Nazionale di Fisica Nucleare, Sezione di Roma 1, $^{aa}$Sapienza Universit\`{a} di Roma, I-00185 Roma, Italy} 

\affiliation{Rutgers University, Piscataway, New Jersey 08855} 
\affiliation{Texas A\&M University, College Station, Texas 77843} 
\affiliation{Istituto Nazionale di Fisica Nucleare Trieste/Udine, $^{bb}$University of Trieste/Udine, Italy} 
\affiliation{University of Tsukuba, Tsukuba, Ibaraki 305, Japan} 
\affiliation{Tufts University, Medford, Massachusetts 02155} 
\affiliation{Waseda University, Tokyo 169, Japan} 
\affiliation{Wayne State University, Detroit, Michigan  48201} 
\affiliation{University of Wisconsin, Madison, Wisconsin 53706} 
\affiliation{Yale University, New Haven, Connecticut 06520} 
\author{T.~Aaltonen}
\affiliation{Division of High Energy Physics, Department of Physics, University of Helsinki and Helsinki Institute of Physics, FIN-00014, Helsinki, Finland}
\author{J.~Adelman}
\affiliation{Enrico Fermi Institute, University of Chicago, Chicago, Illinois 60637}
\author{T.~Akimoto}
\affiliation{University of Tsukuba, Tsukuba, Ibaraki 305, Japan}
\author{M.G.~Albrow}
\affiliation{Fermi National Accelerator Laboratory, Batavia, Illinois 60510}
\author{B.~\'{A}lvarez~Gonz\'{a}lez}
\affiliation{Instituto de Fisica de Cantabria, CSIC-University of Cantabria, 39005 Santander, Spain}
\author{S.~Amerio$^w$}
\affiliation{Istituto Nazionale di Fisica Nucleare, Sezione di Padova-Trento, $^w$University of Padova, I-35131 Padova, Italy} 

\author{D.~Amidei}
\affiliation{University of Michigan, Ann Arbor, Michigan 48109}
\author{A.~Anastassov}
\affiliation{Northwestern University, Evanston, Illinois  60208}
\author{A.~Annovi}
\affiliation{Laboratori Nazionali di Frascati, Istituto Nazionale di Fisica Nucleare, I-00044 Frascati, Italy}
\author{J.~Antos}
\affiliation{Comenius University, 842 48 Bratislava, Slovakia; Institute of Experimental Physics, 040 01 Kosice, Slovakia}
% ...

\collaboration{CDF Collaboration\footnote{With visitors from $^a$University of Massachusetts Amherst, Amherst, Massachusetts 01003,
$^b$Universiteit Antwerpen, B-2610 Antwerp, Belgium, 
$^c$University of Bristol, Bristol BS8 1TL, United Kingdom,
$^d$Chinese Academy of Sciences, Beijing 100864, China, 
$^e$Istituto Nazionale di Fisica Nucleare, Sezione di Cagliari, 09042 Monserrato (Cagliari), Italy,
$^f$University of California Irvine, Irvine, CA  92697, 
$^g$University of California Santa Cruz, Santa Cruz, CA  95064, 
$^h$Cornell University, Ithaca, NY  14853, 
$^i$University of Cyprus, Nicosia CY-1678, Cyprus, 
$^j$University College Dublin, Dublin 4, Ireland,
$^k$Royal Society of Edinburgh/Scottish Executive Support Research Fellow,
$^l$University of Edinburgh, Edinburgh EH9 3JZ, United Kingdom, 
$^m$Universidad Iberoamericana, Mexico D.F., Mexico,
$^n$Queen Mary, University of London, London, E1 4NS, England,
$^o$University of Manchester, Manchester M13 9PL, England, 
$^p$Nagasaki Institute of Applied Science, Nagasaki, Japan, 
$^q$University of Notre Dame, Notre Dame, IN 46556,
$^r$University de Oviedo, E-33007 Oviedo, Spain, 
$^s$Texas Tech University, Lubbock, TX  79409, 
$^t$IFIC(CSIC-Universitat de Valencia), 46071 Valencia, Spain,
$^u$University of Virginia, Charlottesville, VA  22904,
$^{cc}$On leave from J.~Stefan Institute, Ljubljana, Slovenia, 
}}
\noaffiliation

\begin{abstract}
A search for new physics using three-lepton (trilepton) data collected with the CDF II detector and
corresponding to an integrated luminosity of 976 pb$^{-1}$ is presented.
The standard model predicts a low rate of trilepton events, 
which makes some supersymmetric processes,
such as chargino-neutralino production, measurable in this channel.
The $\mu\mu+\ell$ signature is investigated, 
where $\ell$ is an electron or a muon, with the additional requirement of large missing
transverse energy.  In this analysis, the lepton transverse momenta with respect to the 
beam direction ($p_T$) are as low as 5 GeV/$c$, a selection that improves the 
sensitivity to particles which are light as well as to ones which result in leptonically decaying tau leptons.
At the same time, this low-$p_T$ selection presents additional challenges due to the
non-negligible heavy-quark background at low lepton momenta.  
This background is measured with an innovative technique using experimental data.
Several dimuon and trilepton
control regions are investigated, 
and good agreement between experimental results and standard-model predictions is observed.
In the signal region, we observe one three-muon event and expect
$0.4\pm0.1$ $\mu\mu+\ell$ events from standard-model processes.
\end{abstract}

\maketitle
\section{\label{sec:1} Introduction}
The standard model (SM) of particle physics is enormously successful in describing the
known particles and their interactions.  However, strong motivation 
from experimental data as well as important theoretical considerations point to new physics beyond the SM.  
Astrophysical observations that have resulted
in the ``concordance'' model of cosmology \cite{wmap} require a source of dark matter that does not
exist in the SM.   Theoretically, the SM has well-known limitations in explaining the 
origin of mass and solving the hierarchy problem.  Moreover, it does not satisfy our desire for the unification 
of the strong and electroweak interactions and the integration of 
gravity in a unique theory \cite{baertata}. 

\begin{figure}
\centering
\begin{tabular}{cc}
a)&
\begin{minipage}{2in}
\centering
\includegraphics[scale=0.35]
{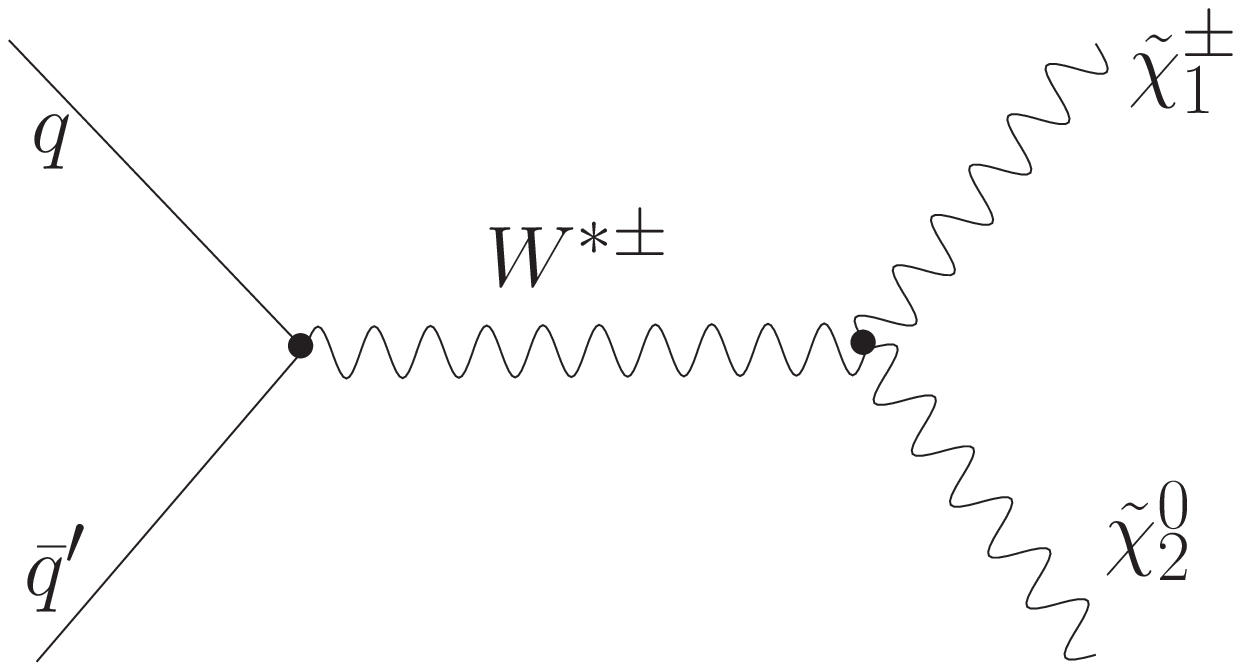}
\end{minipage}
\\
\\
\rr{b)}{-1.5}&
\begin{minipage}{2in}
\centering
\includegraphics[scale=0.35]
{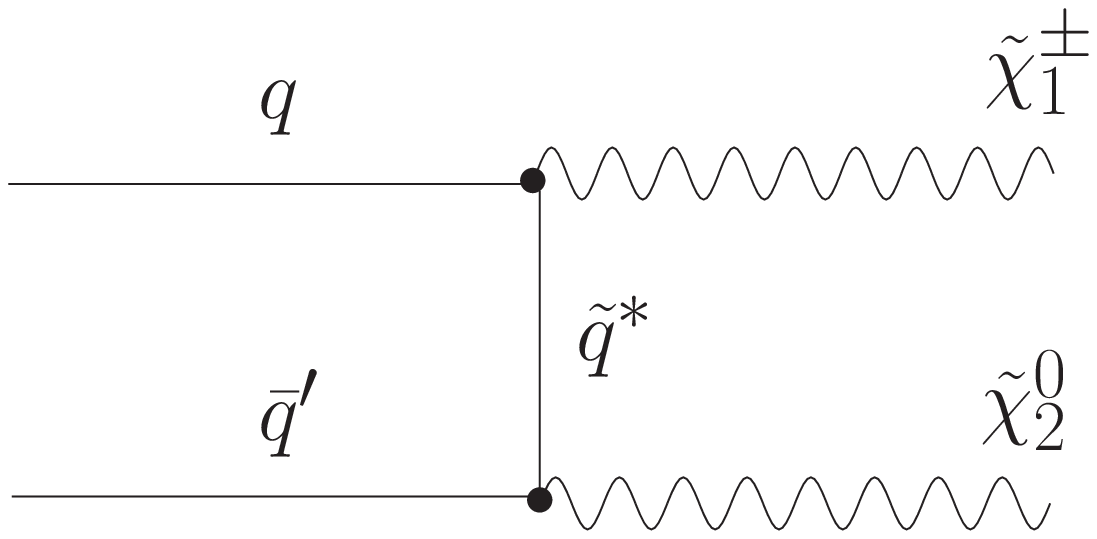}
\end{minipage}
\end{tabular}
\caption{Chargino-neutralino production through an $s$-channel $W$ boson (a) and a $t$-channel squark propagator (b).  
The $t$-channel is suppressed in scenarios with very massive squarks.\label{CHprod}} 
\end{figure}

\begin{figure}[t]
\centering
\begin{tabular}{cc}
\rr{a)}{-1.5} & \rr{c)}{-1.5} \\
\begin{minipage}{1.65in}
\centering
\includegraphics[scale=0.35]
{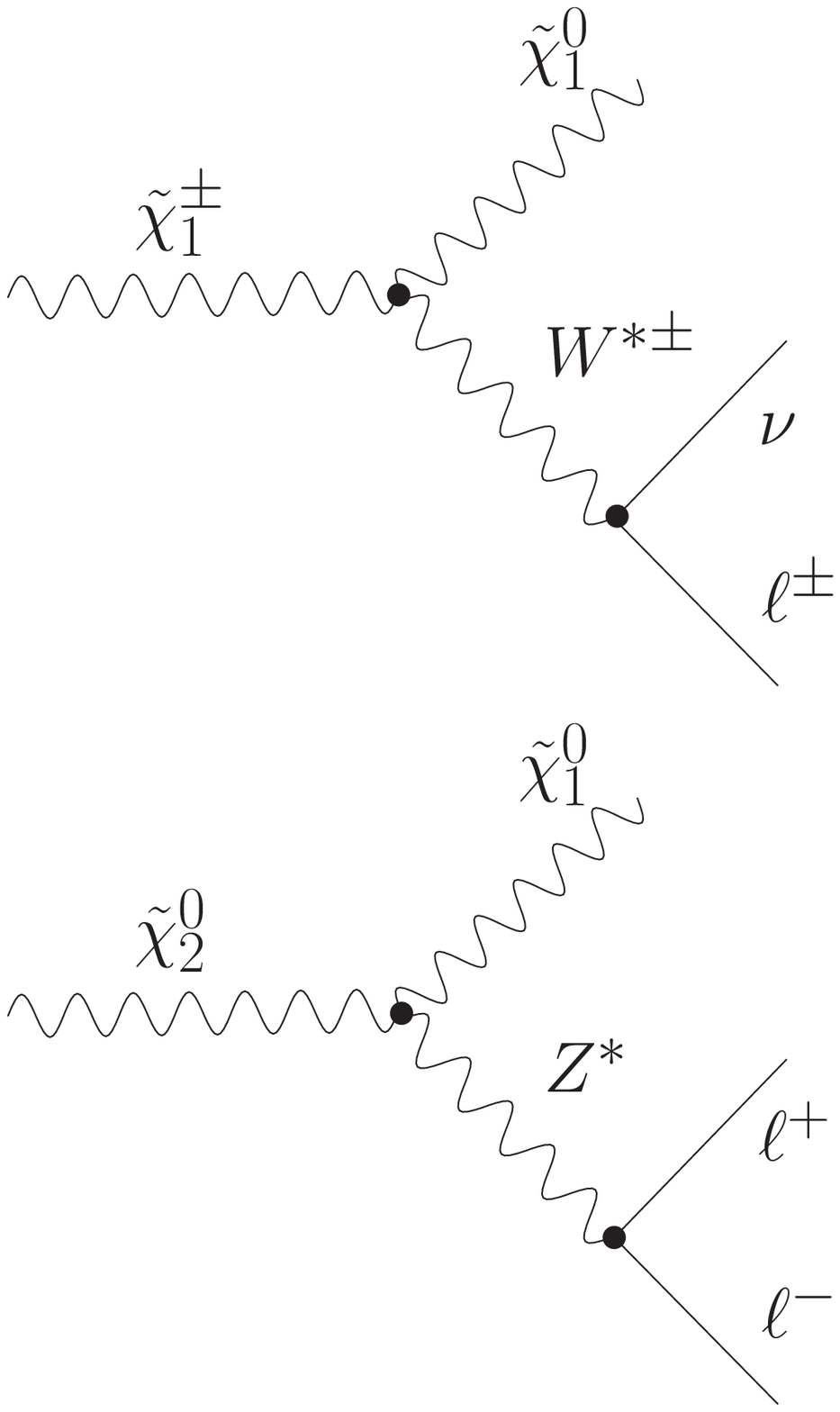}
\end{minipage}
&
\begin{minipage}{1.65in}
\centering
\includegraphics[scale=0.35]
{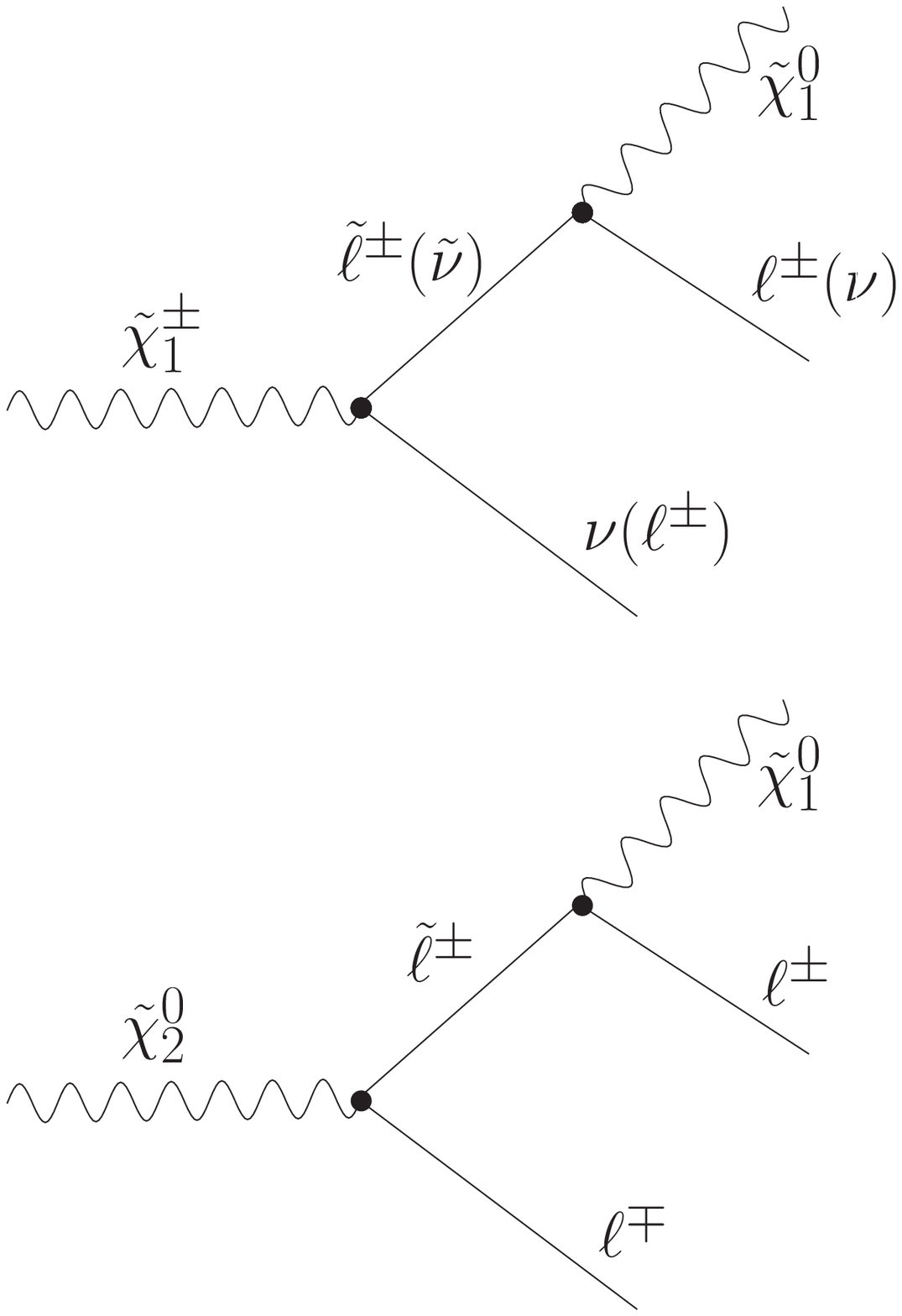}
\end{minipage}\\
\rr{b)}{21} & \rr{d)}{21} \\
\end{tabular}
\caption{Chargino and neutralino decays through gauge bosons (a,b) and through sleptons (c,d).  The branching fractions depend on the masses of the sleptons, which always decay to charged leptons, unlike the gauge bosons.  The leptonic signature consists of three leptons in both cases. \label{CHdecay}} 
\end{figure}

A powerful strategy for discovering new physics 
is to search in event topologies where the SM predicts extremely low production rates. 
One of these topologies is three leptons (trileptons) in hadronic collisions.  
The lepton candidates we observe at the Tevatron collider result mainly from QCD processes or the decays of massive gauge bosons ($W$ or $Z$), or photon conversions.  The leptons rarely appear in multiplicity greater than two. 
The trilepton signature is favored by a large class of models of supersymmety (SUSY) \cite{susy,susy2}, in which the lightest supersymmetric particles are the gauginos, the supersymmetric partners of the gauge bosons.  The corresponding observable SUSY particles are two charginos ($\tilde{\chi}^{\pm}_{1,2}$) and four neutralinos ($\tilde{\chi}^{0}_{1,2,3,4}$), which result from the mixing of the gauginos and the supersymmetric partners of the Higgs bosons, the higgsinos.
The associated production of charginos and neutralinos may have a detectable cross section \cite{footnote1} at the Tevatron and may give rise to trilepton events as shown in Figs.\ \ref{CHprod} and \ref{CHdecay}.  The most common decays are through off-shell vector bosons or scalar leptons (sleptons), with branching fractions that depend on the chargino, neutralino, and slepton masses.  

Under the assumption of $R$-parity \cite{rparity} conservation, SUSY particles cannot yield only SM particles in their decay; the lightest SUSY particle (LSP) will be stable and escape detection.
Therefore, SUSY events would be characterized by large transverse momentum imbalance (``missing transverse energy'', or $\met$).  
In many SUSY scenarios the lightest neutralino is either the LSP or it decays to the LSP resulting in $\met$ in both cases.  Additional $\met$ results from the undetected final-state neutrinos, as shown in Fig.\ \ref{CHdecay}.
The trilepton+$\met$ topology investigated here is the ``golden" signature for the discovery of SUSY at the Tevatron \cite{trileptons1,trileptons2,trileptons3,trileptons4,trileptons5}. 

The LSP is a candidate for the cold dark matter of the universe \cite{darkmatter1,darkmatter2}.  In addition, SUSY offers a solution to the hierarchy problem \cite{hierarchy1,hierarchy2, hierarchy3} and the possibility for unification of interactions at high energies \cite{unification}.  

Searches for chargino and neutralino production have been previously performed by the LEP \cite{lep,lep2} and Tevatron experiments \cite{run1, d0}.  
In this paper we present a trilepton analysis that utilizes increased luminosity and improved kinematic acceptance.  We search for new physics in the final state with two muons and an additional electron or muon using data collected with the CDF II detector from March 2002 to February 2006 from proton-antiproton collisions at $\sqrt{s}=1.96$ TeV.  The integrated luminosity of our sample is 976 ${\rm pb}^{-1}$.  To increase the sensitivity to new light particles and tau leptons that decay leptonically, we use a very low $p_T$ threshold (5 GeV/$c$) for the identified leptons, where $p_T$ is the transverse momentum with respect to the beam direction.  

We define several dimuon and trilepton SM-dominated control regions, in which we verify our understanding of the backgrounds. 
In order to avoid bias, we complete the validation of the background -- both in event yields and kinematic shapes -- in the control regions before investigating the events in the signal region.
Finally, our result is combined with other trilepton searches at CDF \cite{comboPRL} to set a stronger limit on chargino-neutralino production.  Although this search is inspired by SUSY-predicted chargino-neutralino production, the analysis is generic enough to be sensitive to any new physics that would enhance the production of prompt trileptons and $\met$.

This paper is organized as follows.  In Section \ref{sec:2} we describe the CDF II detector.  
In Section \ref{sec:3} we define the experimental dataset and present 
an event selection that reduces the SM background expectation while accepting
events from possible new-physics signals.
Section \ref{sec:4} describes how the SM background rates are estimated,
and Section \ref{sec:5} discusses two SUSY-model scenarios we consider.
In Section \ref{sec:6} we present the determination of the 
systematic uncertainties on the 
signal and background event-yield predictions.  In Section \ref{sec:7} we present the event yields and kinematic distributions in our control regions that increase the confidence in our understanding of the SM background.  Finally, in Section \ref{sec:8} we present the results in the signal region.

\section{\label{sec:2} The CDF II detector}

The CDF II detector \cite{cdfDetector} is a multi-purpose cylindrical detector with projective-tower calorimeter geometry and excellent lepton identification capability.  It operates at the Tevatron collider where protons and antiprotons collide with a center-of-mass energy of 1.96 TeV.  In our coordinate system the positive $z$-axis is defined by the proton beam direction and the positive $y$-axis by the vertical upward direction.  The detector is approximately symmetric in the $\eta$ and $\phi$ coordinates, where the pseudorapidity $\eta$ is defined as $\eta = -\ln(\tan (\theta/2))$, $\theta$ is the polar angle with respect to $\vec{z}$, and $\phi$ is the azimuthal angle.  We briefly present here the CDF components that are most critical to this analysis.

In the center of the apparatus, near the beam collision point,
a silicon detector of inner radius of 1.35 cm and outer radius of 25.6 cm provides detailed tracking in the $|\eta|<2$ region, necessary for the accurate
determination of the proton-antiproton interaction points (primary vertices) and impact parameters of particle trajectories with respect to these points.

A cylindrical 96-layer open-cell argon-ethane (50\%-50\%) drift chamber (COT) of inner radius of 44 cm and outer radius of 132 cm provides tracking for charged particles with $\sim 100\%$ detection
efficiency in the central ($|\eta|<1.1$) region. 
The central tracking system is located in a magnetic field of 1.4 T provided by a superconducting solenoidal magnet.  
The relative resolution in tracking momentum provided by the COT is $\delta p_T/p_T = 0.0017 p_T ({\rm GeV}/c)^{-1}$.  

Surrounding the central tracker, and outside the solenoid, a central electromagnetic calorimeter (CEM) and a central hadronic calorimeter (CHA) measure the energy of electrons, photons, and hadrons.  The CEM is composed of layers of lead and scintillator whereas the CHA is composed of layers of steel and scintillator.
The relative energy resolution is $13.5\%/\sqrt{E_T} \oplus 2\%$ for the CEM and $75\%/\sqrt{E_T} \oplus 3\%$ for the CHA, where the transverse energy $E_T=E\sin\theta$ is quoted in GeV units.
A strip chamber (CES), placed inside the electromagnetic calorimeter at the position of maximum development of the electromagnetic shower (six radiation lengths), is used for shower shape determination and for matching the calorimeter energy depositions with COT tracks.  In the forward region, a plug electromagnetic calorimeter  ($1.1<|\eta|<2.4$) has a relative resolution of $16\%/\sqrt{E_T} \oplus 0.7\%$ and a plug hadronic calorimeter ($1.3<|\eta|<2.4$) a resolution of  $130\%/\sqrt{E_T} \oplus 4\%$.
The raw missing transverse energy vector is defined as $-(\sum_i \vec{E}_{iT})$, where $\vec{E}_{iT}$ has magnitude equal to the energy deposited in the $i^{\rm th}$ calorimeter tower and direction perpendicular to the beam axis and pointing to that calorimeter tower.

Outside the calorimeters, the central muon system consists of drift chambers.
The central muon chambers (CMU) detect muons in the pseudorapidity range $|\eta|<0.6$, while the central muon extension (CMX) chambers detect muons in the $0.6<|\eta|<1.0$ range, both with a detection efficiency of almost $100\%$ for muons above 3 GeV/$c$.  To reduce the hadron punch-through contamination, extra chambers (CMP) are installed outside the CMU chambers, with extra steel absorber added between them.
The muons that are detected by both CMU and CMP chambers are labeled ``CMUP muons'', and their detector signatures cannot be easily caused by hadrons.

The instantaneous luminosity is measured with Cherenkov counters located close to the beam line at $3.7 < |\eta| <4.7$.

The CDF trigger system \cite{uiuc} has a three-level pipelined and buffered architecture; each level provides 
a rate reduction sufficient to allow for processing at the next level with minimal deadtime.  The first level 
consists of special-purpose processors that accept events at rate of 25 kHz, with an average
event size of 170 kB, counts main triggering objects and feeds the second level with an event rate 
of 350 Hz.  The second level is also based on hardware and performs a partial event reconstruction
before passing the events to the next level.  Finally a software-based third level uses a fast version 
of the offline event reconstruction to reduce the event rate to 75 Hz, appropriate for writing 
to tape.  The track-based triggers account for approximately 75\% of the trigger bandwidth and are 
used in this analysis.  For a muon trigger, the main requirement is that a COT track is geometrically matched 
to a track segment in a muon detector.
 
\section{\label{sec:3} The CDF Dataset and signal-region event selection}

In order to include in our analysis muons and electrons that come from tau decays, 
we use a low transverse momentum requirement ($p_T>5$ GeV/$c$) for these leptons. 
For this reason we analyze data collected with the CDF low-$p_T$ dimuon triggers ($p_T$ nominally above 4 GeV/$c$ for both muons).  
These muons are central in the detector (CMUP or CMX).  We measure the trigger efficiency 
using $J/\psi$, $\Upsilon$ and $Z$-boson events collected with single-muon triggers.  In these samples, we remove hadronic backgrounds using the mass-spectra sidebands, and count the frequency that a second muon fired the trigger of interest.
The plateau value of the 
trigger efficiency's $p_T$ dependence for single muons is $\sim 0.95$ and it is reached at $p_T \sim 5$ GeV/$c$.

After the collected events are processed by the offline reconstruction software, additional requirements are applied for the definition of the dimuon sample.  
We require that each event has a primary vertex within 60 cm from the nominal center 
of the detector in the $z$ direction and that at least two muons with transverse momenta above 5 GeV/$c$ originate from that primary vertex and pass the CDF standard muon tracking and calorimetry requirements and track-chamber matching requirements \cite{generic}.  
In events with more than one reconstructed primary vertex,
we use the primary vertex that is closest to the tracks of the two highest-$p_T$ muons that satisfy all other event requirements.  We specifically
require that two good-quality COT tracks are geometrically matched with respective reconstructed track segments in the CMX or CMU+CMP detectors, 
that the energies deposited in the electromagnetic and hadronic calorimeters are consistent with that expected from minimum ionizing particles, 
and that the two muons are isolated.  We define the isolation $I$ as the energy
deposited in the calorimeters in a cone of $\Delta R = \sqrt{(\Delta \phi)^2 + (\Delta \eta)^2}=0.4$ 
around the muon without counting the energy deposited by the muon.
We require that $I<0.1\times p_T c$ if $p_T>20$~GeV/$c$ or $I< 2$~GeV otherwise, where $p_T$ is the transverse momentum of the muon.
The selected two muons are also $\Delta R >0.4$ apart.
A critical requirement is that the muons are prompt as measured by the impact parameter ($d_0$), 
defined as the distance of closest approach of a track to the primary vertex in the transverse plane.  
We require that $|d_0|<0.02$ cm if the 
the muon leaves tracking signals in the silicon detector (silicon hits)
and that $|d_0|<0.2$ cm if the muon leaves no silicon hits.
We expect that most
muons with large impact parameters come from heavy flavor (bottom- or charm-hadron semileptonic decays), fake muons (light-flavor hadrons such as pions and kaons that decay in flight or punch-through to the muon detectors), and cosmic rays. 
The heavy flavor (HF) and fake-muon backgrounds dominate at low dimuon masses.  
Residual cosmic-ray background, not removed by the cosmic filters described in \cite{generic}, is reduced
by requiring that the three-dimensional angular separation ($\Delta \varphi$) of the two highest-$p_T$ muons is less than 178 degrees.
After including 
the selection criteria discussed above, the total muon identification 
efficiency, as measured with $J/\psi$ and $Z$ boson CDF data, is (90-96)\%,
rising with increasing muon $p_T$.

For the trilepton selection, we require the presence of a third muon satisfying the same selection requirements
as the first two, or an electron satisfying the CDF standard electron calorimeter, tracking,  and track-calorimeter matching identification 
requirements \cite{generic}. 
The transverse energy and momentum of an electron 
is required to exceed 5 GeV.  Tracks associated with electrons
should match hits in the CES wires.  
We require that $I<0.1\times E_T$ if 
$E_T>20$~GeV or $I<2$~GeV otherwise, where $I$ is now the energy-based
isolation of the electron, and $E_T$ is its transverse energy.
Electrons originating from photons that convert into $e^+e^-$ pairs are 
identified with an algorithm \cite{generic} that 
seeks nearby tracks with a common vertex and direction.
These electrons are removed from the observed data sample. 
The electron identification efficiency is (75-83)\% \cite{generic},
rising with increasing electron transverse energy, as measured with Drell-Yan \cite{dy} electrons.  The third lepton is required to be $\Delta R>0.4$ away from the leading two muons. 
 
We define the signal region by the following additional
requirements: the dimuon mass (constructed using the two highest-$p_T$ muons) is greater than 15 GeV/$c^2$,
for removal of low-mass resonances, and outside a $Z$ mass window of $76<M_{\mu\mu}<106$~GeV$/c^2$.
In addition, we require the missing transverse energy ($\met$) to exceed 15 GeV,
in order to select events with undetected new particles while rejecting Drell-Yan, HF, 
and fake-muon backgrounds.
Finally, we count the number of jets $N_{\rm jets}$ with energy above 15 GeV and
we require that $N_{\rm jets}\le 1$, in order to reduce the $t\bar{t}$ background.  
In this analysis we use jets defined by a fixed-cone algorithm \cite{generic}
with a cone size of $\Delta R=0.4$.  We require that jets deposit less than 90\%
of their measured energy in the electromagnetic calorimeter,
in order to avoid counting electrons or photons as jets.  Jet energies
are corrected \cite{jes} to represent better the energy of the final-state hadrons.
Global and local corrections are applied as well as inclusion of
corrections for the effects of multiple interactions.
These corrections are also applied to the raw missing transverse energy
for the calculation of $\met$, which is also corrected for the 
presence of muons in our events. 
We check the consistency of the observed data compared to the SM predictions
in the control regions that are described in Section \ref{sec:7}.

\section{\label{sec:4} Standard-Model Backgrounds}

To determine the significance of any incompatibility between prediction and observation, and also to set limits on production cross sections and masses of new particles, we need a
reliable background estimation.
The major SM source of dimuons is the Drell-Yan (DY) process and, in events with low dimuon mass,
HF production and the fake-lepton background.  
In the trilepton regions, the dominant backgrounds are DY (accompanied by
a fake lepton), dibosons ($WW$, $ZZ$, and $WZ$), and HF.  Because HF and fake leptons are difficult to model 
with Monte Carlo (MC) simulations due to sizable higher-order QCD effects and the imperfect modeling of the lepton isolation in a high particle-multiplicity hadronic environment, we estimate these backgrounds using CDF data.
All other backgrounds are estimated with MC simulation.

\subsection{MC-estimated backgrounds}

We use the {\sc pythia} \cite{pythia} generator to model the DY, $WW$, $ZZ$, and $t\bar{t}$ background, and
{\sc madevent} \cite{madevent} for the $WZ$ background \cite{footnote2}.
The DY background includes the decays to tau leptons that subsequently decay to muons \cite{tauola}.
We use the {\sc cteq5l} \cite{cteq} parton distribution functions (PDF) throughout.
For the trilepton predictions we require the reconstructed electrons and muons to be kinematically matched with the generator-level leptons, in order not to double-count some of the fake-lepton contribution.  
To estimate the trilepton background from
DY+$\gamma$, we relax this matching requirement, demand that the electron is identified at the event-simulation level
as a photon-conversion product, and normalize the surviving event using a scale factor \cite{anadiPRD}.  This scale factor accounts
for the difference in conversion-removal inefficiency between the observed data and the MC simulation.
In the remainder of the paper we add the DY+$\gamma$ background to the rest of the diboson contribution
($WW$, $ZZ$, and $WZ$).
We process each generated event with the CDF detector simulation, based on {\sc geant} \cite{geant}.
We normalize all samples using the leading-order theoretical cross sections 
multiplied by the appropriate scale (``K-factor'') to correct for next-to-leading order effects
\cite{dibosonCross, ttCross}.  Scale factors that correct for the known differences in lepton 
identification and reconstruction efficiencies between the observed data and the MC simulation are also applied.  
\subsection{Data-estimated backgrounds}

We first estimate the fake-lepton background, using an independent CDF data sample.
Subsequently, we use this fake-lepton background and the MC-estimated DY contribution
in our HF-estimation method.

\subsubsection{Fake leptons \label{fakeleptons}}
``Fake'' leptons are reconstructed lepton candidates that are either not real leptons or are real leptons but are neither  prompt nor do they originate from semileptonic decays of HF quarks.  In the case of muons, the fakes can be light-flavored hadrons, such as pions and kaons or part of hadronic showers, that penetrate (``punch through'') the calorimeters and reach the muon detectors or decay to muons in flight.  In the case of electrons, fakes are jets that are misreconstructed as electrons, often due to neutral pions that decay to photons, which shower in the electromagnetic calorimeter.  We can thus associate the fake leptons with light-flavor partons.  Using multijet CDF datasets collected with jet-based triggers, we measure the ``fake rate'', {\it i.e.}, 
the probability for an isolated track to be misreconstructed as a muon or the probability for 
a jet to be misreconstructed as an electron.    
The fake-rate is measured as a function of the track's (jet's) transverse momentum (energy) and pseudorapidity.  The fake rate is of the order of $10^{-2}$ for isolated tracks to be incorrectly reconstructed as muons and increases with the $p_T$ of the muon candidate's track.  The fake rate of a jet being reconstructed as an electron is of the order of $10^{-4}$ and falls with increasing $E_T$.  The fake rates increase for higher pseudorapidity leptons \cite{anadiPRD}.

To determine the background coming from a real muon and a misidentified hadron (i.e., fake dimuon background), we use single-muon low-$p_T$-triggered CDF data.  For each event, we require one good muon candidate that passes the requirements of our analysis.  We then apply the fake rate on all other tracks in the event, 
except on the track of a second muon (to remove DY contamination of the fake background). 
We remove events in which the ``muon+track'' mass is within the $Z$ boson window ($76<M_{\mu\mu}<106$~GeV$/c^2$) and also $35<\met<55$ GeV.
These events are associated with decays of real $Z$ bosons produced at rest, where one decay muon is not detected, resulting in $\met$ equal to about half the mass of the $Z$ boson.  We investigate the heavy flavor contamination in this light-flavor-dominated background, caused by a real muon coming from a heavy-quark semileptonic decay, which is misreconstructed as an isolated track instead of a muon.  This contamination is negligible (approximately 0.2\% of the background), mainly because of the background-rejection power of our muon isolation requirement.  Because the single-muon low-$p_T$ trigger was not always present during data taking, the dimuon fakes extracted using this trigger are normalized to the default dimuon CDF data luminosity, in order to represent the size of fake dimuon contamination in our analysis CDF dataset.  In CDF data with no $\met$ or jet multiplicity cuts applied (``inclusive'' dataset), $\sim (9 \pm 5)\%$ of the dimuons are fake.  In the signal region, the dimuon fake contamination is $\sim (16 \pm 8)\%$.

For the determination of the background coming from a real muon pair and a misidentified hadron (i.e., fake trilepton background), we use our dimuon low-$p_T$-triggered CDF dataset, require two good muons, and model the fake third lepton by applying the fake rate to the extra tracks.  We assume that the number of events with two fake leptons is negligible, given the low value of the fake rates. 
In order not to over-count the trilepton fakes, we require the three leptons in our signal MC and background MC samples to be kinematically matched with the generated ones.  
The fake trilepton background is determined to be $\sim (50\pm 25)\%$ of the total background in both the inclusive dataset and the signal region.

\subsubsection{Heavy flavor \label{heavyflavor}}
\begin{figure*}[!]
\includegraphics[scale=.7]{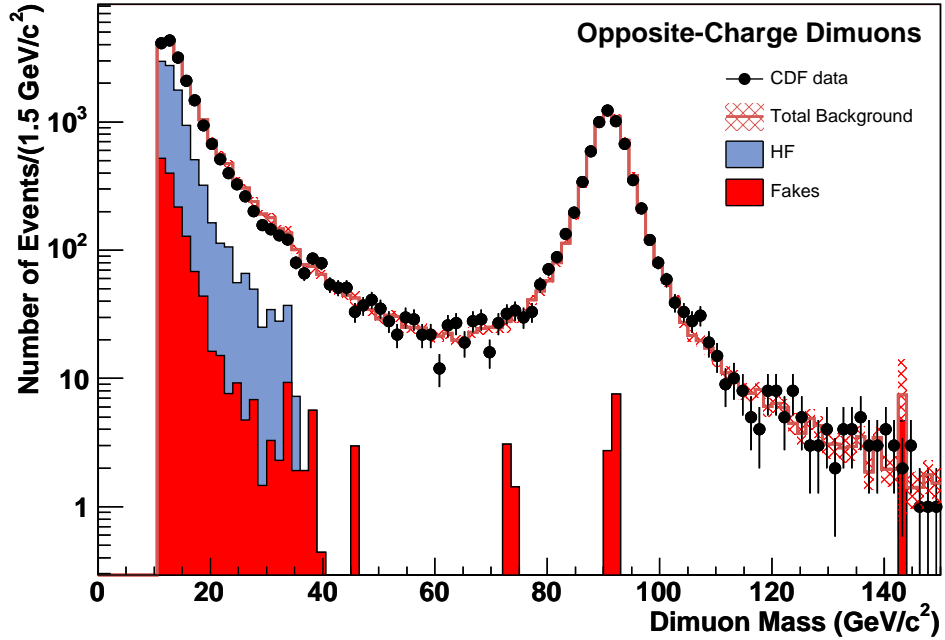}
\caption{Fit of HF+DY+fakes dimuon mass distribution to the observed data for opposite-charge dimuons.  The HF normalization is the only free parameter of the fit.  The blue (light gray) filled histogram is the HF and the red (dark gray) is the fakes.  The thick line represents the total background, which is almost exclusively DY, HF, and fakes at the opposite-charge dimuon level.  The hatched areas indicate the total uncertainty on the prediction.}
\label{os}
\end{figure*}
\begin{figure*}[!]
\includegraphics[scale=.7]{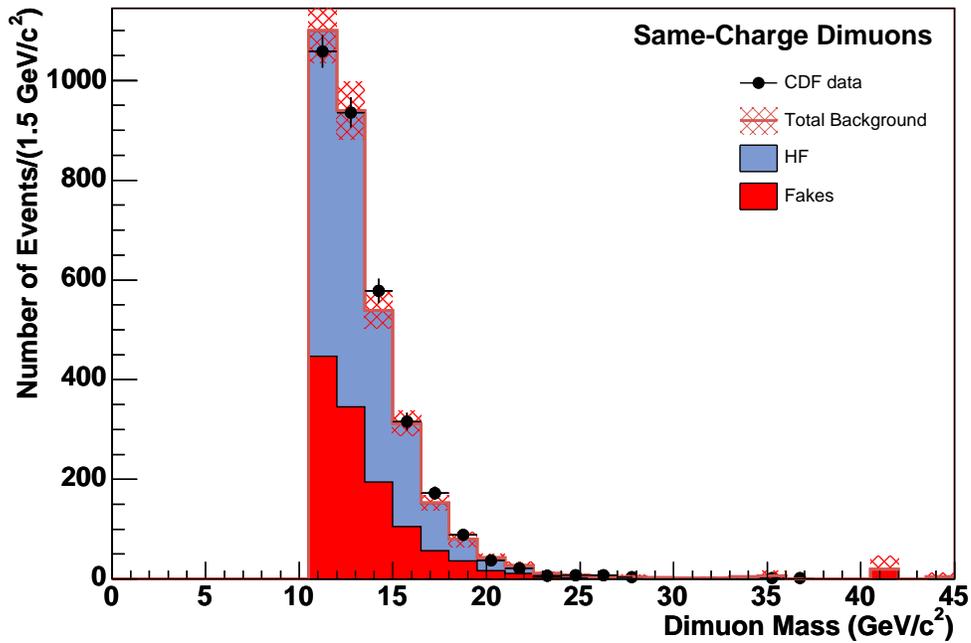}
\caption{Fit of HF+DY+fakes dimuon mass distribution to the observed data for same-charge dimuons.  The HF normalization is the only free parameter of the fit.  The blue (light gray) filled histogram is the HF and the red (dark gray) is the fakes.  The thick line represents the total background, which is constituted almost exclusively by HF and fakes for same-charge dimuon pairs.  The hatched areas indicate the total uncertainty on the prediction.}
\label{ss}
\end{figure*}

One of the most significant challenges of this analysis is the consideration of muons with 
transverse momentum as low as 5 GeV/$c$.  
This low $p_T$ requirement increases our acceptance,
but at the same time
contaminates our sample with HF and fake-lepton events.

We present here an innovative technique for the determination of the amount of this background using the observed data.  
We construct an HF-rich (HFR) CDF dataset by reversing the impact parameter requirement for at least 
one of the observed muons, so that the absolute value of the muon impact parameter is above 0.02 cm if 
there are silicon hits associated with the muon track, or above 0.2 cm if 
there are no silicon hits.
We also require the dimuon mass to be less than 
35 GeV/$c^2$.  Monte Carlo studies show that above that value we expect mainly DY
and a negligible HF background.  We investigated the expected dimuon mass spectrum of DY and fake-lepton 
in the HFR sample and we determined that the effect of the contamination is negligible.

We subsequently use the HFR dimuon mass shape 
combined with the absolute fake dimuon mass distribution plus the absolute DY 
dimuon mass distribution from MC simulation in order
to fit the observed data.  All data samples other than HFR include 
the low impact parameter requirement.  
Because we observe negligible DY in the same-charge dimuon channel,
we perform the fit for same-charge and opposite-charge dimuons separately.
This helps us validate our HF-estimation method in the HF-rich same-charge
dimuon environment.
The only free parameter of the fits is the HF normalization -- the DY contribution is fixed based 
on the theoretical cross section and the integrated luminosity of the observed data, and the fake-lepton background
distribution is fixed based on the absolute expectation, as described in Section \ref{fakeleptons}.  The results of the fits can be seen 
in Figs.\ \ref{os} and \ref{ss}, for opposite-charge and same-charge muons, respectively.  From 
the two fits, we extract two HF normalization factors that are applied as weights to the original 
unweighted same-charge and opposite-charge dimuon HFR events, in order to describe the HF background in 
the observed data.  The same weights are used in all kinematic control regions.
The weight for the opposite-charge HF dimuons is $1.94 \pm 0.04$ and for the same-charge HF dimuons is $1.12 \pm 0.05$, where the uncertainties come from the fits.  This tells us that we expect almost twice as many opposite-charge HF events in the low-impact parameter region compared to the high-impact parameter one.  
Overall, the ratio of opposite-charge HF to same-charge HF events in the inclusive observed data after the normalization is $\sim$ 4:1, a value that is also verified with $b\bar{b}$/$c\bar{c}$ MC simulation and is a result of the conserved charge in the underlying quark-pair production and the rates of cascade semileptonic decays of $b$-hadrons and $c$-hadrons.
In regions with no HFR events, we estimate the size of this background
by extrapolating the HF prediction from neighboring dimuon control regions that contain sufficient numbers of events.

The trilepton HF background is estimated by requiring that the normalized HFR sample has a third lepton.
If there are no events satisfying this requirement, then we extrapolate from either neighboring dimuon or
trilepton control regions with sufficient statistics.
For example, we have no HFR data in the trilepton signal region.  We estimate the HF background there by extrapolating from the low-$\met$ region, 
where we have trilepton HFR events.  For the extrapolation we use the dimuon $\met$ distribution, 
using the fact the $\met$ distribution is similar for dimuon and trilepton events.  We verify this fact with 
the use of MC-simulated $b\bar{b}$/$c\bar{c}$ events.

For the determination of the systematic uncertainty associated with the HF-estimation method,
we re-estimate the HF background by redefining the HFR dataset using either a requirement on the number
of silicon detector hits for the muon that has large impact parameter (at least two silicon hits), and/or applying a requirement on the impact
parameter significance ($|d_0|/\delta|d_0|>5$).  These cuts favor HF events but reduce our HFR dataset statistics.
The HF-estimation method systematic uncertainty is about 25\% in the signal region.  

Although the HF normalization is extracted from the inclusive analysis sample, with dimuon mass greater than 10.5 GeV/$c^2$ (to avoid the $\Upsilon$ resonances) and no additional $\met$ or jet multiplicity requirements, the agreement of our HF predictions in both event yields and kinematic 
distributions for all our dimuon and trilepton control regions is excellent, as we show in Section \ref{sec:7}.

\section{\label{sec:5} SUSY Signal Scenarios}

This analysis is a generic search for trilepton events in which we focus
on minimizing the SM background.
We nevertheless consider two mSUGRA \cite{msugra} SUSY signal scenarios, ``SIG1'' and ``SIG2'', defined
by the value of the common sfermion mass ($m_0$) and common gaugino
mass ($m_{\frac{1}{2}}$) at unification scale, the trilinear coupling ($A_0$),
the ratio of the two Higgs fields vacuum expectation values ($\tan\beta$), and the sign of the higgsino mixing parameter (${\rm sign}(\mu)$): 

\begin{itemize}

\item SIG1: $m_0$=100 GeV/$c^2$, $m_{\frac{1}{2}}$=180 GeV/$c^2$, $A_0$=0, $\tan\beta$=5, $\mu > 0$.
The expected cross section $\sigma(p\bar{p} \rightarrow \tilde{\chi}^{\pm}_{1} \tilde{\chi}^{0}_{2})$ times the branching ratio $\cal B$ to leptons is $\sigma \times {\cal B} = 0.642 \times 0.22$ pb.
The cross section was obtained using the next-to-leading order calculation of {\sc prospino} \cite{prospinoChargino} and the branching ratio using {\sc pythia}.
The corresponding chargino and lightest neutralino masses would be 116 GeV/$c^2$ and 65 GeV/$c^2$, respectively.

\item SIG2: $m_0$=74 GeV/$c^2$, $m_{\frac{1}{2}}$=168 GeV/$c^2$, $A_0$=0, $\tan\beta$=3, $\mu > 0$.
The expected cross section times the branching ratio to leptons is $\sigma \times {\cal B} = 1.023 \times 0.5$ pb, as
given by {\sc prospino} and {\sc pythia}.
The corresponding chargino and lightest neutralino masses would be 103 GeV/$c^2$ and 57 GeV/$c^2$, respectively.

\end{itemize}

These two signal scenarios serve as benchmarks of possible SUSY signal and were used for the optimization of the 
minimum $\met$ requirement in the signal region, which is set at 15 GeV \cite{footnote3}.  The mass spectrum of the supersymmetric particles was obtained with {\sc isajet} \cite{isajet} and the events are generated with {\sc pythia}.  
The SIG1 scenario leads to three-body decays (Fig.\ 2a,b) of the lightest chargino (${\tilde{\chi}}_1^{\pm}$) and the next-to-lightest neutralino (${\tilde{\chi}}_2^{0}$) with branching ratios to
electrons and muons suppressed, due to the low branching ratio of the gauge bosons to leptons.  On the other hand, the SIG2 scenario leads exclusively to two-body decays (Fig.\ 2c,d) of
both gauginos to sleptons, with ${\tilde{\chi}}_1^{\pm}$ decaying always to a final-state tau lepton (produced from a stau decay).
Our analysis is more sensitive to SIG2, due to the higher cross section and our ability to select events with low momentum final-state leptons, originating from tau decays.

\section{\label{sec:6} Systematic Uncertainties}

The sensitivity of our search to signals of new physics and 
the significance of a potential excess of events are influenced by the 
uncertainties on our background estimates.
Because we perform a counting experiment, we concentrate on the 
uncertainties on the expected number of background and signal events.
The event-yield systematic uncertainty is naturally different 
for MC-simulated and CDF-data-estimated physical processes.  
We first discuss the systematic uncertainty on the MC-estimated 
backgrounds and SUSY signals and then we treat the systematic
uncertainty on the CDF-data-based background from HF and fake leptons.

The sources of systematic uncertainty in the signal region,
with their effect on signal and MC-estimated background event yields are:

\begin{itemize}

\item the luminosity uncertainty (6\%) \cite{lumi,lumi2},

\item the lepton-identification scale factors uncertainty ($\sim 10\%$), 

\item the trigger efficiency uncertainty ($\sim 1\%$),

\item the jet-energy scale uncertainty ($\sim 1\%$) \cite{jes};  
this source of systematic uncertainty is responsible for migrating
events from one control or signal region to another, since variations in 
jet energies affect both the corrections to the $\met$ and the jet 
multiplicity,

\item the PDF uncertainty (1\%-2\%) \cite{cteq}.

\item the uncertainty from the theoretical cross-sections estimates (5-12\% depending on the process) \cite{dibosonCross, ttCross},

\item the uncertainty on the initial- and final-state QCD-induced radiation (ISR/FSR) \cite{isr}, which has an effect of 
4\% and 12\% for background and signal MC samples, respectively, and 

\item the uncertainty induced from the limited MC statistics: for the SIG2 MC it is $\sim 2\%$ for the dimuons and 
$\sim 6\%$ for the trileptons; for the standard-model background MC it is $\sim 3\%$ for the dimuons and
$\sim 40\%$ for the trileptons (the latter mainly due to the DY+$\gamma$ limited MC statistics).

\end{itemize}

All of the above sources of systematic uncertainty are correlated among the different
physics processes (DY, diboson, $t\bar{t}$), 
with the exceptions of the cross section systematic uncertainties and the MC samples' statistical uncertainties.  
Still, the
sources of systematic uncertainties are uncorrelated with each other and the respective uncertainties
are summed in quadrature with each other to give the total yield uncertainties in the control and signal regions.

The systematic uncertainty associated with the HF-estimation method
consists of a part that is anti-correlated with the DY+fakes systematic uncertainty (because the HF weights are given by the fit of the DY+fakes+HF to the observed data, and a varied level
of DY+fakes affects these weights) and an uncorrelated part (from the fit uncertainty of about 2-4\%
for the fixed DY+fakes level and from the HF-estimation method systematic uncertainty, as described in Section \ref{heavyflavor}).  The correlated DY+HF+fakes systematic uncertainty is about 22\%.
The fake-lepton uncertainty is set to a conservative maximum-envelope 50\% level, which is determined by studying different jet-triggered CDF samples \cite{anadiPRD}.
For the total systematic uncertainty of the predicted SM event yield in all control regions we take into
account all correlations among physics processes.  For each source of systematic uncertainty affecting the MC samples, we vary all MC samples (including
DY) in a correlated manner and redo the fit of DY+fakes+HF to the observed data to extract a new HF estimation.  The total
variation gives us the total effect of the systematic uncertainty.  The same procedure is followed when we include
the fake-lepton estimation uncertainty and its effect on HF due to the above fit.

The effect of the systematic uncertainties on the background and SUSY signal expected event yields in the signal region 
can be found in Tables \ref{syst1} and \ref{syst2} respectively.
\begin{table}[t]
\centering
\caption{The dimuon and trilepton event-yield systematic uncertainties for the backgrounds in the signal region.  The uncertainties are summed based on the contributions of the separate backgrounds taking into account all correlations.  The upper part of the table shows the MC-related systematic uncertainties whereas the lower part shows the systematic uncertainties for the CDF-data-estimated backgrounds.  The uncertainties due to the MC statistics are not shown. \label{syst1}}
\begin{tabular}{cll}\hline \hline
\multicolumn{1}{c}{Source} & 
\multicolumn{1}{c}{Dimuons} &
\multicolumn{1}{c}{Trileptons}\\ \hline 
Electron scale factors & -- & $\pm$2\% \\
Muon scale factors & $\pm$8\% & $\pm$5\% \\
Luminosity & $\pm$3\% & $\pm$2\% \\
Trigger efficiency & $\pm$0.5\% & $\pm$0.2\% \\
PDF & $\pm$2\% & $\pm$1\% \\
ISR/FSR & $\pm$2\% & $\pm$1\% \\
Theoretical cross sections & $\pm$3\% & $\pm$2\% \\
Jet-energy scale & $\pm$0.5\% & $\pm$0.02\% \\
\hline
{\bf Total MC syst.} & $\pm$9\% & $\pm$6\% \\
Fakes estimation & $\pm$8\% & $\pm$25\% \\
HF estimation & $\pm$5\% & $\pm$2\% \\
\hline \hline
{\bf Total (with correlations)} & $\pm$10\% & $\pm$24\% \\
\hline \hline
\end{tabular}
\end{table}
\begin{table}[t]
\centering
\caption{The dimuon and trilepton event-yield systematic uncertainties in the signal region, for the SIG1 and SIG2 SUSY scenarios.}
\begin{tabular}{cll}\hline \hline
\multicolumn{1}{c}{Source} & 
\multicolumn{1}{c}{Dimuons} &
\multicolumn{1}{c}{Trileptons}\\ \hline 
Electron scale factors & -- & $\pm$6\% \\
Muon scale factors & $\pm$11\% & $\pm$15\% \\
Luminosity & $\pm$6\% & $\pm$6\% \\
Trigger efficiency & $\pm$0.9\% & $\pm$0.5\% \\
PDF & $\pm$1\% & $\pm$1\% \\
ISR & $\pm$12\% & $\pm$12\% \\
Theoretical cross sections & $\pm$10\% & $\pm$10\% \\
Jet-energy scale & $\pm$0.3\% & $\pm$0.6\% \\
MC statistics & $\pm$2\% & $\pm$6\% \\
\hline \hline
{\bf Total} & $\pm$20\% & $\pm$24\% \\
\hline \hline
\end{tabular}
\label{syst2}
\end{table}
\begin{figure*}[!]
\includegraphics[scale=.7]{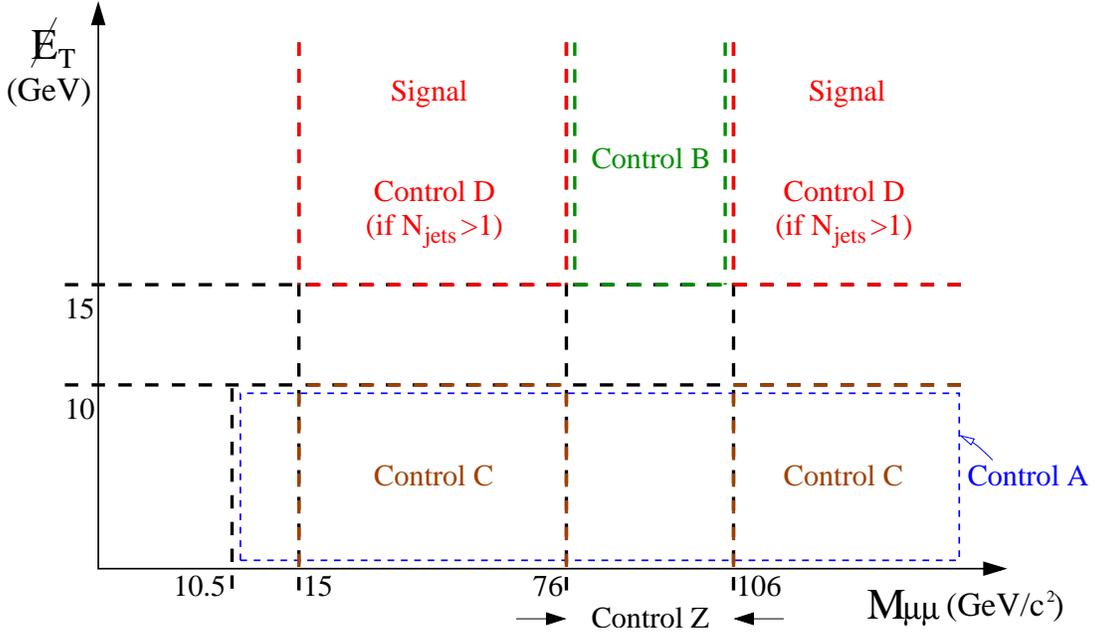}
\caption{The control and signal regions used in our analysis are defined in the
dimuon mass vs.\ $\met$ plane, with the extra requirement of low ($\leq 1$) or high ($> 1$)
jet multiplicity. In this paper we show results for the control regions that result 
from the inversion of one of the three main kinematic selections (dimuon mass, missing transverse energy, 
and jet multiplicity), with the addition of a $Z$-mass control region (Control {\sc z}) and a low-$\met$  
control region (Control {\sc a}).  The control regions above are defined for low jet multiplicity, unless otherwise
stated.\label{controlRegionsPlot}}
\end{figure*}
\begin{table*}[!]
\footnotesize
\center
\caption{Expected and observed dimuon event yields, in all control regions and the signal region.  The expected SUSY signal event yield
is for the SIG2 mSUGRA scenario.  Combined statistical and systematic uncertainties are shown and correlations among sources of systematic uncertainty are included.  The signal region without a requirement for a third lepton is a dimuon control region.}
\begin{tabular}{ccccccccc}\hline \hline
\multicolumn{1}{c}{Region} & 
\multicolumn{1}{c}{DY} &
\multicolumn{1}{c}{HF} & 
\multicolumn{1}{c}{Fakes} &
\multicolumn{1}{c}{Diboson} & 
\multicolumn{1}{c}{$t\bar{t}$} &
\multicolumn{1}{c}{\bf Total SM expected} &
\multicolumn{1}{c}{\bf SUSY expected} &
\multicolumn{1}{c}{\bf Observed}\\ \hline 
Control {\sc z} & $6419 \pm 709$ & - & $10\pm 11$ & $2.4 \pm 0.2$ & $1.18\pm 0.14$ & $6433\pm 712$ & $0.30 \pm 0.07$ &$6347$ \\
Control {\sc a} & $14820\pm 2242$ & $9344\pm 1612$ & $2294\pm 1148$ & $1.03 \pm 0.09$ & $0.12\pm 0.03$ & $26459\pm 1429$ & $0.9 \pm 0.2$& $26295$ \\
Control {\sc b} & $217\pm 25$ & -  & $9\pm 7$ & $1.7 \pm 0.2$ & $0.27\pm 0.05$ & $227\pm 26$ & $0.5 \pm 0.1$& $253$ \\
Control {\sc c} & $5770\pm 1043$ & $2238\pm 384$ & $466\pm 234$ & $0.49 \pm 0.07$ & $0.02\pm 0.01$ & $8474\pm 857$ & $0.7 \pm 0.2$& $8205$ \\
Control {\sc d} & $7.8\pm 1.5$ & $9\pm 4$ & $0.3\pm 0.3$ & $0.21 \pm 0.07$ & $4.1\pm 0.4$ & $22\pm 5$ & $1.8 \pm 0.4$& $23$ \\
Signal Reg.& $169\pm 30$ & $90\pm 20$ & $49\pm 25$ & $6.5 \pm 0.4$ & $0.96\pm 0.11$ & $315\pm 37$ & $17 \pm 3$& $297$ \\
\hline \hline
\end{tabular}	
\label{crDil}
\end{table*}
\begin{table*}[!]
\footnotesize
\center
\caption{Expected and observed trilepton event yields, in all control regions and the signal region.  The expected SUSY signal event yield
is for the SIG2 mSUGRA scenario.  Combined statistical and systematic uncertainties are shown and correlations among sources of systematic uncertainty are included.}
\begin{tabular}{ccccccccc}\hline \hline
\multicolumn{1}{c}{Region} & 
\multicolumn{1}{c}{DY} &
\multicolumn{1}{c}{HF} & 
\multicolumn{1}{c}{Fakes} &
\multicolumn{1}{c}{Diboson} & 
\multicolumn{1}{c}{$t\bar{t}$} &
\multicolumn{1}{c}{\bf Total SM expected} &
\multicolumn{1}{c}{\bf SUSY expected} &
\multicolumn{1}{c}{\bf Observed}\\ \hline 
Control {\sc z} & $0.2\pm 0.2$ & -  & $ 2.5\pm 1.2$ & $0.26 \pm 0.06$ & - & $3 \pm 1$ & $0.06 \pm 0.01$ & $4$ \\
Control {\sc a} & $0.3\pm 0.2$ & $6\pm 3$ & $7.6\pm 3.8$ & $0.25 \pm 0.08$ & - & $14\pm 4$ & $0.08 \pm 0.02$ & $16$ \\
Control {\sc b} & - & - & $0.2\pm 0.1$ & $0.094 \pm 0.009$ & - & $0.3\pm 0.1$ & $0.10 \pm 0.03$ & $0$ \\
Control {\sc c} & $0.2\pm 0.2$ & $3\pm 2$ & $2\pm 1$ & $0.10 \pm 0.06$ & - & $5 \pm 2$ & $0.06 \pm 0.02$ & $8$ \\
Control {\sc d} & - & - & $0.02 \pm 0.01$ & $0.003 \pm 0.002$ & $0.011 \pm 0.008$ & $0.03\pm 0.01$ & $0.04 \pm 0.02$ & $0$ \\
{\bf Signal Reg.} & {\bf -} & {\boldmath $0.06\pm 0.04$} & {\boldmath $0.2\pm 0.1$} & {\boldmath $0.15 \pm 0.06$} & {\bf -} & {\boldmath $0.4\pm 0.1$} & {\boldmath $1.7 \pm 0.4$} & {\boldmath $1$} \\
\hline \hline
\end{tabular}	
\label{crTri}
\end{table*}
 \begin{figure*}[t!]
\includegraphics[scale=0.95]{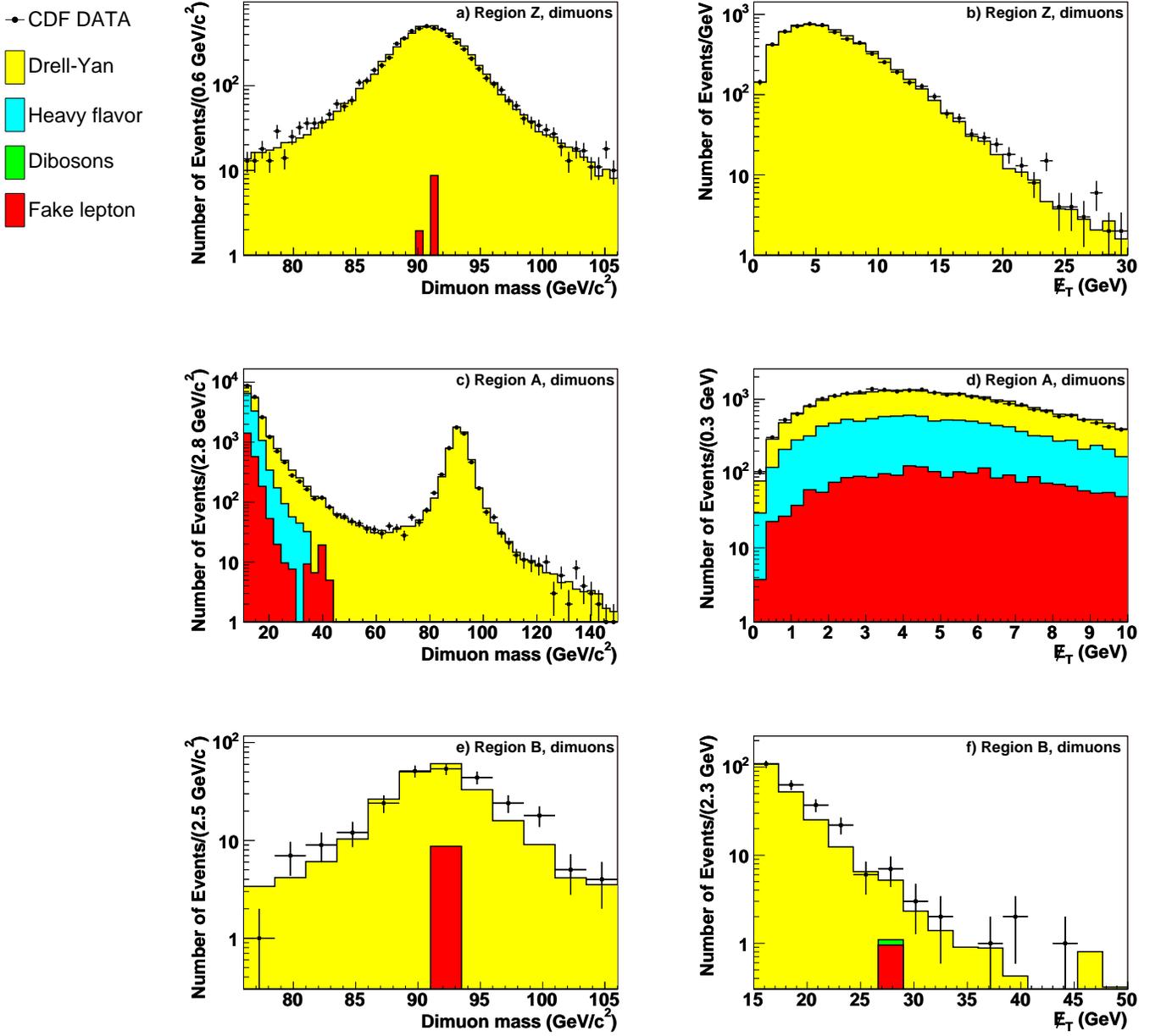}
\caption{Dimuon mass and $\met$ distributions for the SM background in the dimuon control regions {\sc z} (a,b), {\sc a} (c,d), and {\sc b} (e,f).  The background histograms are stacked.  The CDF data are indicated by points with error bars.\label{app1}}
\end{figure*}
\begin{figure*}[!]
\includegraphics[scale=0.95]{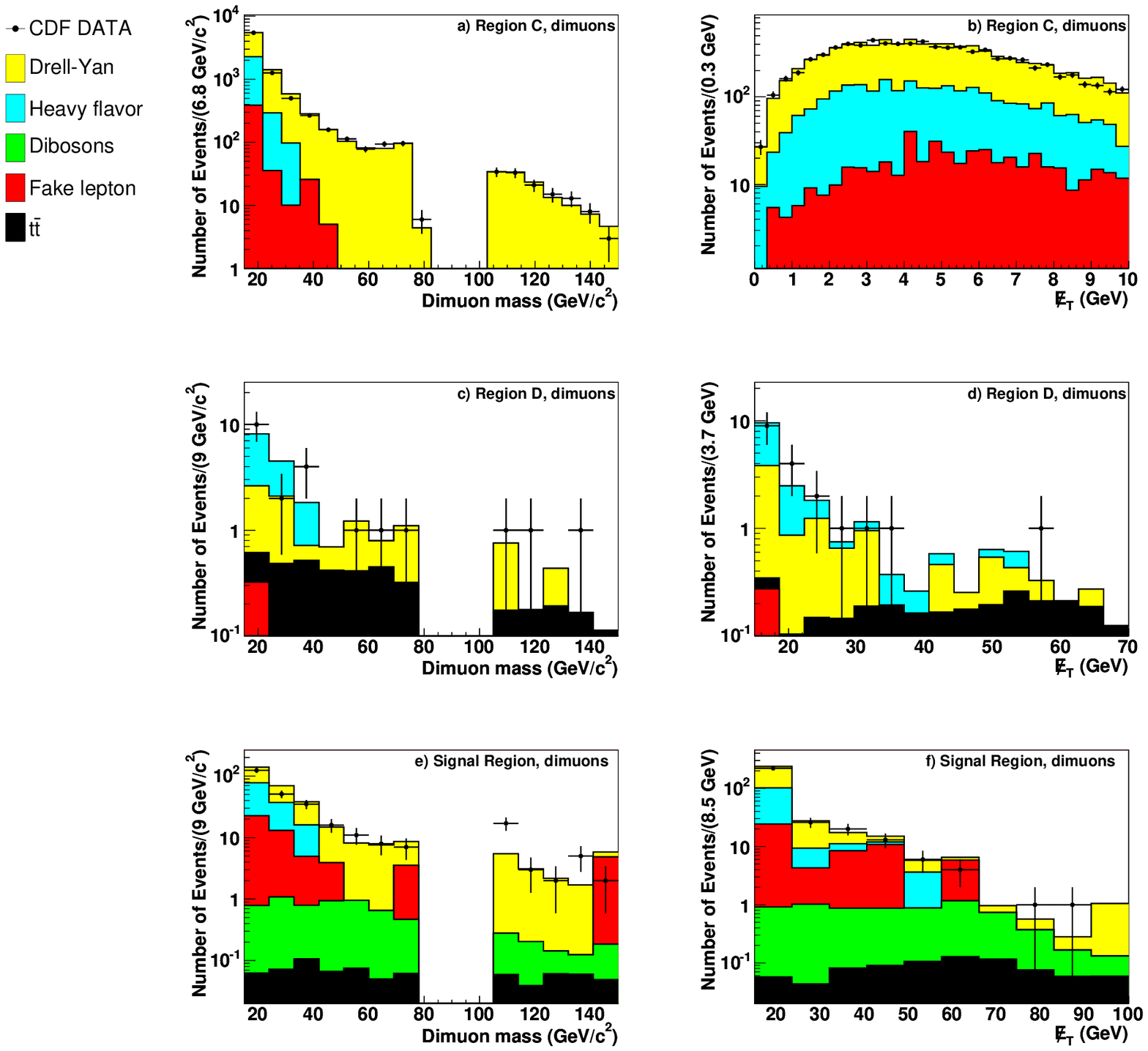}
\caption{Dimuon mass and $\met$ for the SM background in the dimuon control regions {\sc c} (a,b), {\sc d} (c,d), and dimuon signal region (e,f). The background histograms are stacked.  The CDF data are indicated by points with error bars.\label{app2}}
\end{figure*}
\begin{figure*}[!]
\includegraphics[scale=0.94]{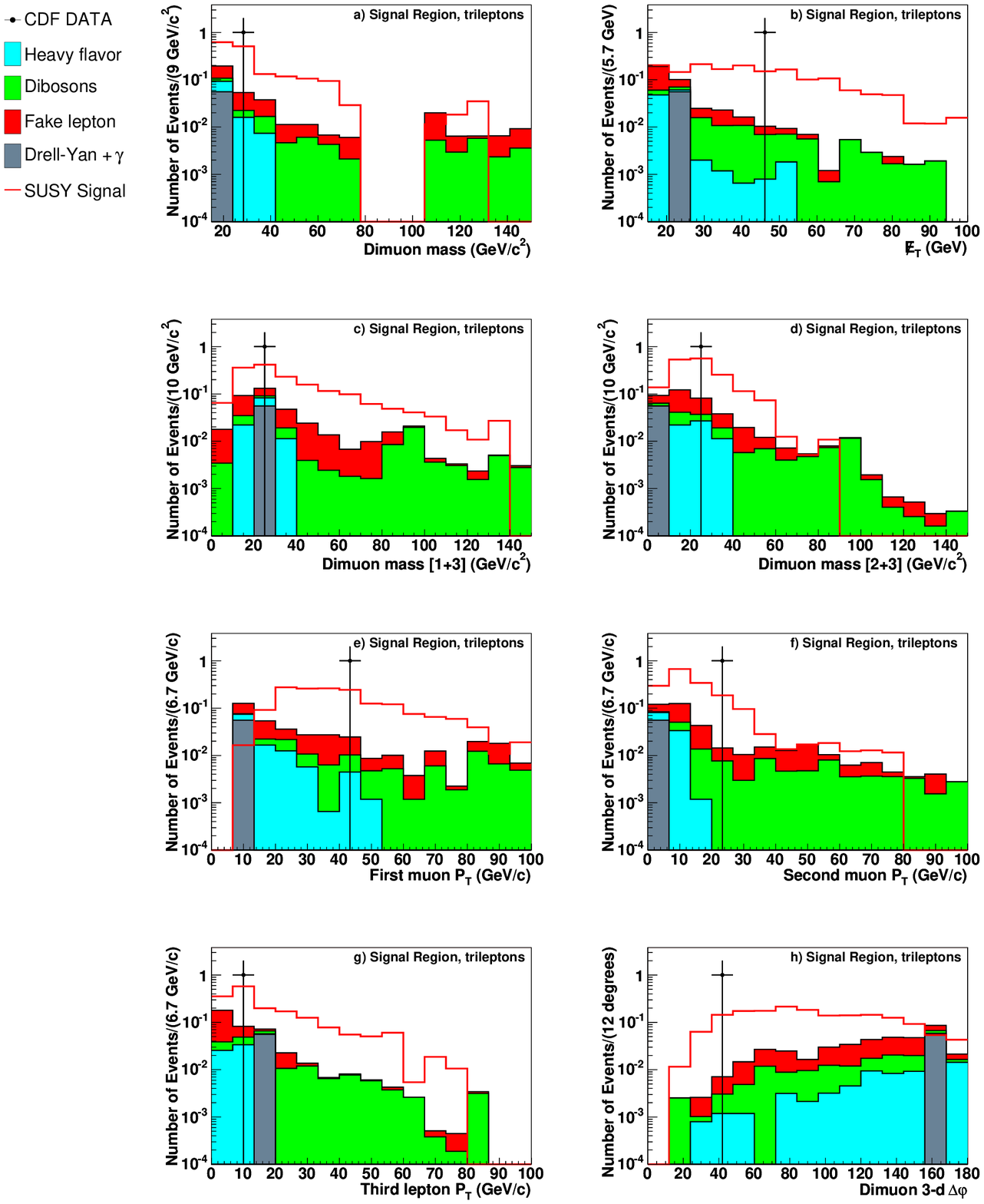}
\caption{Kinematic variables for the SM background and the SIG2 SUSY signal, in the trilepton signal region.  The background histograms are stacked; the signal histogram is not.  The CDF data are indicated by points with error bars.\label{app3}}
\end{figure*}
\section{\label{sec:7} Control Regions}

We investigate control regions 
defined by the dimuon mass, $\met$, and jet multiplicity,
as shown in Fig.\ \ref{controlRegionsPlot}. Overall, 22 dimuon and trilepton control regions are defined.  Most control regions we investigate are naturally SM-dominated with 
little expectation of SUSY signal.  Low $\met$ and $M_{\mu\mu}$ regions are dominated by the HF background,
whereas the $76 <M_{\mu\mu} <106$ GeV/$c^2$ region is almost exclusively populated with $Z$ bosons.
The 5 GeV gap in the $\met$ cuts between the signal and the control
regions ensures that the low $\met$ control regions contain a negligible
amount of signal.
We compare the SM event-yield predictions with observed events 
in the control regions (along with kinematic plots) before looking at the signal region.  

We present here a $Z$ boson resonance control region (``Control {\sc z}''), a low $\met$ control region (``Control {\sc a}''), and three control regions (``Control {\sc b}, {\sc c}, and {\sc d}'') that result from the inversion of one of the three signal region requirements at a time [dimuon mass ($M_{\mu\mu}$), or $\met$, or jet multiplicity ($N_{\rm jets}$) respectively].
In region {\sc z}, we require that the muons have opposite charge and that the dimuon mass lie between 76 and 106 GeV/$c^2$.
We use this control region for validating the luminosity, the trigger efficiencies, and muon-identification scale factors.
In region {\sc a}, we require that $\met<10$ GeV and $M_{\mu\mu}>10.5$ GeV/$c^2$.  This region is used for verifying our knowledge of HF and fake-lepton backgrounds.
In region {\sc b}, we require $\met>15$ GeV, dimuon mass within the $Z$-mass region, and low jet multiplicity (at most one jet).  This region helps us verify our background
prediction in a low-yield region as most $Z$ events are characterized by low $\met$.
In region {\sc c}, we require $\met<10$ GeV, exclusion of the $Z$ mass region, $M_{\mu\mu}>15$ GeV/$c^2$, and low jet multiplicity.  This region along with region {\sc a} are the ones with the highest population of HF events.
In region {\sc d}, we require $\met>15$ GeV, exclusion of the $Z$ mass region, $M_{\mu\mu}>15$ GeV/$c^2$, and high jet multiplicity (more than one jet).  This region is expected to be the most sensitive to $t\bar{t}$ production.
We finally study the dimuon events with all signal-region kinematic cuts applied, but before the requirement for a third lepton.  This is a critical control region as the trilepton signal is a subset of this region. 

Table \ref{crDil} shows the expected and observed number of dimuon events in our control regions, and Table \ref{crTri} shows the expected and observed number of trilepton events.  After requiring the presence of a third electron or muon, only control regions 
{\sc z}, {\sc a} and {\sc c} are populated with experimental data.  Region {\sc z} trilepton event yields establish our understanding of the electron fakes, since the third electron in $Z$ boson events is almost exclusively a non-prompt electron.  On the other hand, trilepton regions {\sc z} and {\sc c} confirm our understanding of the HF and fake backgrounds for trileptons. 
\begin{figure*}[t]
\centering
\begin{tabular}{cc}
\hspace{-0.3cm}
\begin{minipage}{3.4in}
\centering
\includegraphics[scale=0.7]{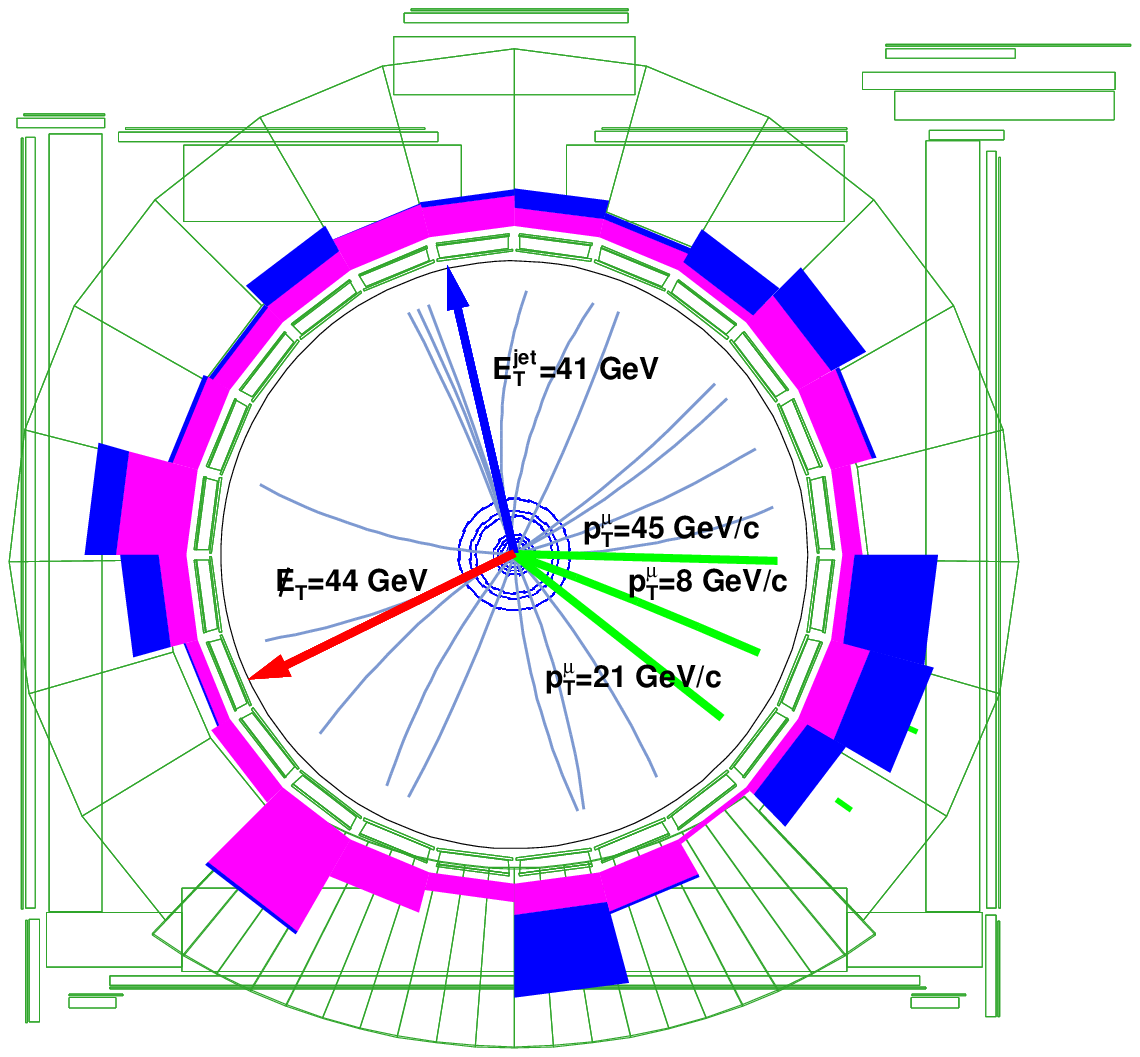}
\caption{The trimuon event in the transverse view of the central CDF detector.  Tracks with transverse momenta above 1~GeV/$c$ are shown.\label{cot1}}
\end{minipage}\hspace{0.6cm}
&
\begin{minipage}{3.37in}
\centering
\includegraphics[scale=0.5]{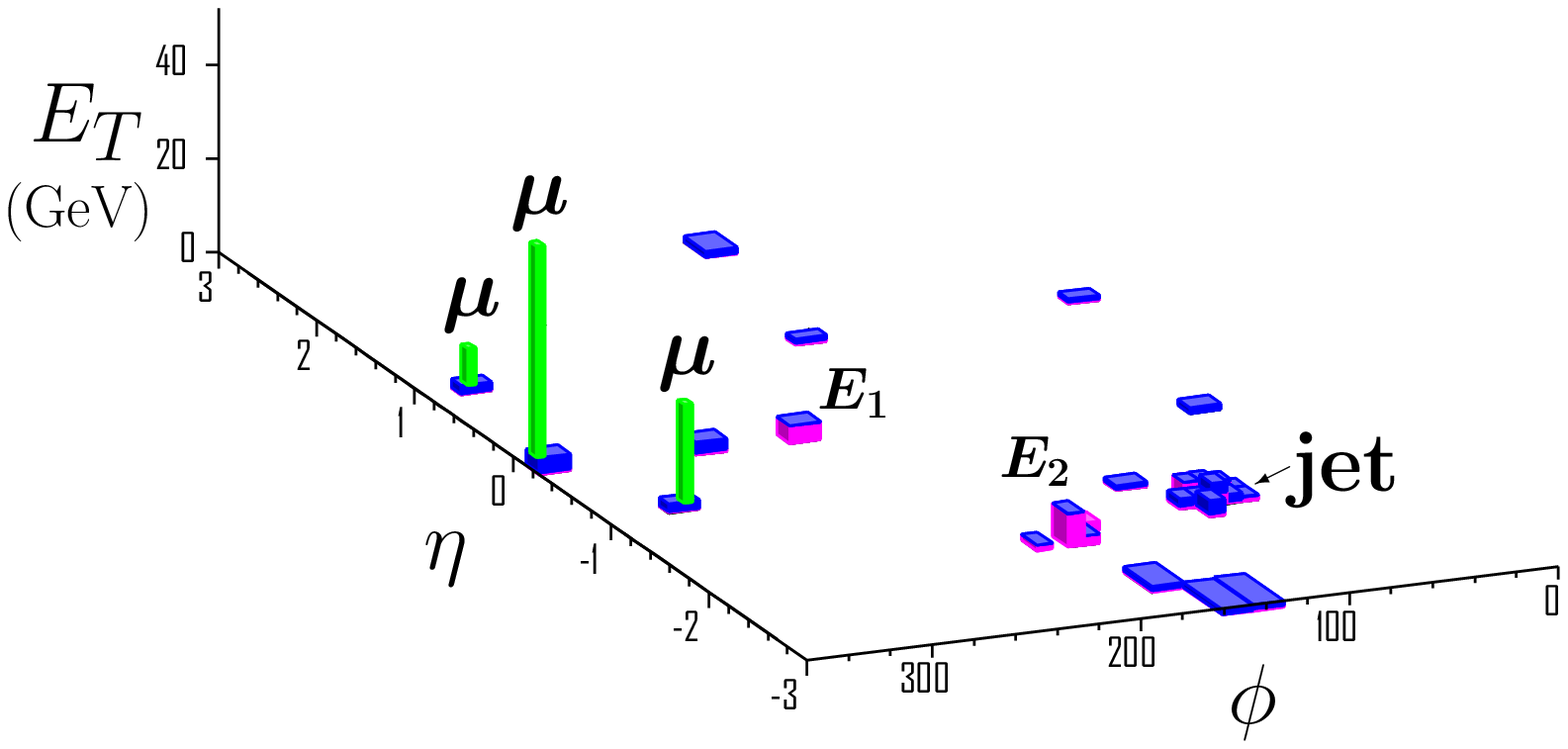}
\caption{The trimuon event in the $\eta-\phi$ view.  Calorimeter transverse energies above 1 GeV are shown.  The longer bars correspond to the track momenta of the three muons (high $\phi$).
The calorimeter energy depositions $E_1$ and $E_2$ are mainly electromagnetic and could be associated with two photons.\label{lego}}
\end{minipage}
\end{tabular}
\end{figure*}
\begin{table*}[!]
\vspace{-0.3cm}
\footnotesize
\center
\caption{Observed trimuon event properties.}
\begin{tabular}{cc}\hline \hline
Kind of muons & CMUP - CMX - CMX\\
$p_T$ of muons (GeV/$c$) & 45.0, 21.1, 7.8\\
$\eta$ of muons & -0.2, -0.9, 0.8\\
$\phi$ of muons (deg.) & 359, 321, 340 \\
Isolation of muons (GeV)& 2.4, 0.2, 1.1\\
Charge of muons & -1, 1, -1\\
Dimuon masses (GeV/$c^2$) & 29.3(1\&2), 21.7(1\&3), 25.7(2\&3)\\
Transverse mass (muon+$\met$) & 86.4, 51.4, 34.2\\
3-d $\Delta\varphi$(leading muons) (deg.) & 46.3 \\
$\met$(GeV) & 43.8\\
$\met$ $\phi$ (deg.) & 205.6\\
Number of Jets & 1\\
$E_T$ of jet (GeV) & 41.1\\
$\eta$ of jet & -1.6\\
$\phi$ of jet (deg.) & 102.9\\
\hline \hline
\end{tabular}
\label{Tevent}
\end{table*}
Figs.\ \ref{app1} and \ref{app2} show the dimuon mass and $\met$ distributions for the dimuon control regions.  The agreement between observed data and prediction in the control regions is satisfactory, both in event yields and kinematic distributions.

\section{\label{sec:8} Signal Region Result}

After observing satisfactory agreement between experimental data and SM predictions in both the dimuon and trilepton control regions,
we look at the CDF data in the trilepton signal region.  We observe one event containing three muons (trimuon event).  
The event is characterized by low track activity and three well-identified muons that
are produced within $\sim 40$ degrees in $\phi$.  Two of the muons are energetic, with transverse momenta of 
45 and 21 GeV/$c$, and the third one is a soft muon with $p_T$ of 8 GeV/$c$. 
Table \ref{Tevent} shows the main properties of this event.  
It is interesting to note the close values of all three dimuon masses.
The event includes one hadronic jet and two energy clusters of mostly electromagnetic energy with transverse energy of $\sim 41$, $\sim 9$, and $\sim 4$ GeV respectively.  The jet and the muons originate from the same and only high-quality primary vertex.  If the electromagnetic energy clusters correspond to real photons, the event would also be interesting in the gauge-mediated supersymmetry breaking (GMSB \cite{gmsb}) scenario, where the lightest neutralino decays to a photon and a gravitino, which is the LSP.  In that case, the final leptonic signature of the chargino-neutralino production would be three leptons, two photons, and $\met$.

Fig.\ \ref{app3} shows where the one trimuon event observed in the signal region appears in the expected distributions of kinematic variables for the signal and the backgrounds.
Kinematic distributions include the three-dimensional opening angle between the leading muons, $\Delta\varphi$.  Figs.\ \ref{cot1} and \ref{lego} show the transverse and lego detector displays, respectively, for this trimuon event.  The Poisson probability to see one event or more, when we expect $0.4\pm0.1$, is 32.6\%. 

It is interesting also to interpret this event in the context of a search only
for trimuon events.  The diboson backgrounds remain, but the large source
of fakes in the dimuon+$e$ sample is reduced.  The total trimuon background
estimation is $0.16\pm0.04$ events.
The Poisson probability to observe one event or more, when we expect $0.16\pm0.04$, is 14.7\%.  We conclude that our event yield is statistically consistent with the SM prediction, noting that most of the kinematics of the event - especially the three-dimensional opening angle of the leading muons $\Delta\varphi$ - are consistent with new physics expectation, as can be seen in Fig.\ \ref{app3}. 

We have combined the results of this analysis with other CDF trilepton analyses to set exclusion limits in several models.  For mSUGRA with no slepton mixing, we set a lower limit for the chargino mass of 129 GeV/$c^2$, which corresponds to an upper limit in $\sigma \times {\cal B}$ of about 0.25~pb at the 95\% confidence level \cite{comboPRL}.

\section{Acknowledgments}
We thank the Fermilab staff and the technical staffs of the participating institutions for their vital contributions. This work was supported by the U.S. Department of Energy and National Science Foundation; the Italian Istituto Nazionale di Fisica Nucleare; the Ministry of Education, Culture, Sports, Science and Technology of Japan; the Natural Sciences and Engineering Research Council of Canada; the National Science Council of the Republic of China; the Swiss National Science Foundation; the A\ P\ Sloan Foundation; the Bundesministerium f\"ur Bildung und Forschung, Germany; the Korean Science and Engineering Foundation and the Korean Research Foundation; the Science and Technology Facilities Council and the Royal Society, UK; the Institut National de Physique Nucleaire et Physique des Particules/CNRS; the Russian Foundation for Basic Research; the Ministerio de Ciencia e Innovaci\'{o}n, and Programa Consolider-Ingenio 2010, Spain; the Slovak R\&D Agency; and the Academy of Finland.


\begin{thebibliography}{42}
\bibitem{footnote1}
\bibinfo{note}{Chargino-neutralino production cross-sections times branching
  ratios to leptons of the order of 0.1 fb have not been excluded yet}.
\bibitem{rparity}
\bibinfo{note}{The $R$-parity is defined as $(-1)^{3(B-L)+2S}$, where $B$ is
  the baryon number, $L$ is the lepton number and $S$ is the spin of the
  particle.}
\bibitem{footnote2}
\bibinfo{note}{{\sc pythia} does not include the $W\gamma$ diagrams when it
  generates $WZ$.}
\bibitem{footnote3}
\bibinfo{note}{The minimum $\met$ threshold is the one that maximizes the ${\rm
  Signal}/\sqrt{{\rm Background}}$ quantity.}
\end{thebibliography}
\end{document}